\documentclass[prl, twocolumn, aps, amssymb, amsfonts]{revtex4}

\begin{document}

\title{Quantification of  quantum correlation of ensemble of states}

\author{Micha{\l} Horodecki, Aditi Sen(De), and Ujjwal Sen}
\affiliation{Institute of Theoretical 
Physics and Astrophysics, University of Gda{\' n}sk, 80-952 Gda{\' n}sk, Poland}

\begin{abstract}

We present first measure of quantum correlation of an ensemble of multiparty states. It is based on the idea of 
minimal entropy production in a locally distinguishable basis measurement. It is shown to be 
a relative entropy distance from a set of ensembles. For bipartite ensembles, which span the 
whole bipartite Hilbert space, the measure is bounded below by average relative entropy of entanglement.
We naturally obtain a monotonicity axiom for any measure of 
quantum correlation of ensembles. We evaluate this 
measure for certain cases. Subsequently we use this measure to propose a complementarity relation between 
our measure and the accessible information obtainable about the 
ensemble under local operations. The measure along with the monotonicity axiom
are well-defined even for the case of a single system, where 
the complementarity relation is seen to be yet another face of the ``Heisenberg uncertainty relation''.

\end{abstract}

\maketitle

\def\com#1{{\tt [\hskip.5cm #1 \hskip.5cm ]}}

Quantifying quantum correlation of states is a deeply studied problem \cite{IBMhuge, VPRK,
 Vidal-mon2000, Horodecki_limits, qic} in 
quantum information. Several measures has been proposed. Prominent among them are 
distillable entanglement \cite{IBM_purification95,IBMhuge}, 
entanglement of formation \cite{IBMhuge}, relative entropy of entanglement \cite{VPRK}, etc.

However to our knowledge, the problem of quantification of quantum correlation of an 
\emph{ensemble} of states has never been studied.
In this paper we propose first measure of quantum correlation (\({\cal Q}\)) of an ensemble  of states. 
A trivial way to quantify quantum correlation of an ensemble of states is to take an average 
of an individual property of the constituent states in the ensemble. E.g., one could take the average 
entanglement of formation \cite{IBMhuge}
of the ensemble states. However such an averaging over a property of the individual states
cannot 
capture the complexity of the ensemble as a whole. To take into account the structure of the ensemble as a whole,
the measure must depend on the ensemble as a whole. 
A simple measure of quantum correlation of multiparty ensembles could be the difference \(\mathbb{D}\) between 
the globally accessible information and the locally accessible information. 
However accessible information is an operationally useful quantity, and one aim of defining 
\({\cal Q}\) is to estimate accessible information under different sets of operations.
Moreover \(\mathbb{D}\) vanishes
 for single-party ensembles, whereas \({\cal Q}\) will be seen to be nontrivial also in such cases.

After defining the measure, we show that it can be 
seen as a relative entropy distance from a set of ``classical'' ensembles. For a bipartite 
ensemble, which spans the whole bipartite Hilbert space, we show that measure is bounded below by 
the average relative entropy of entanglement \cite{VPRK}.
In discussing some of its properties, we naturally obtain 
a monotonicity axiom for any measure of quantum correlation of ensembles.
In a natural way, this gives us the first axiom for 
any measure of quantumness of an ensemble of a single system. We then evaluate the measure for the case of four 
Bell states (given with equal prior probabilites). We also evaluate the measure for  
other more general ensembles.
We subsequently propose a complementarity relation 
\cite{comp1,comp} between this measure of quantum correlation of an ensemble
and the accessible information obtainable about the ensemble, under the relevant class of allowable operations.
In the case of a single system, this complementarity is partially proven, and is seen to be
yet another face
of the ``Heisenberg uncertainty relation''. Importantly, both the information gathering and 
disturbance terms in this relation are information-theoretic, in contrast to e.g. \cite{FuchsPeresEnglert}. 

Consider an  ensemble of bipartite states 
\({\cal E} = \{p_x, \left|\psi_x\right\rangle^{AB}\}_{x=1}^{N}\), 
that are 
not necessarily orthogonal. One can for example think of a source that 
produces the state \(\left|\psi_x\right\rangle^{AB}\) (belonging to the Hilbert space
\({\cal H}_A \otimes {\cal H}_B\)) with probability \(p_x\) (\(x = 1, 2, \ldots, N\)), 
and sends it to two far apart parties Alice(A) and Bob(B). Here 
of course the probabilities \(p_x\) are nonnegative and sum up to unity. 
 Let \({\cal A}\) be the span 
of the states \(\left|\psi_x\right\rangle^{AB}\). 
Let \(\{\left|a_i\right\rangle^{AB}\}_{i=1}^{n}\) be an \emph{orthonormal} basis of \({\cal H}_A \otimes {\cal H}_B\).
Consider the part of  \(\{\left|a_i\right\rangle^{AB}\}_{i=1}^{n}\) which has a 
nonzero overlap with \({\cal A}\) \cite{overlap}. 
This part (with nonzero overlap with \({\cal A}\)) can be distinguishable or indistinguishable under local 
operations and classical communication (LOCC). 
Let  \({\cal D}_{\cal A}\) be the set of 
all such  bases \(\{\left|a_i\right\rangle^{AB}\}_{i=1}^{n}\) whose overlap with \({\cal A}\) is 
LOCC-distinguishable. 
Of course, \(n = \dim \{ {\cal H}_A \otimes {\cal H}_B \}
\leq N\).

The measure of quantum correlation \({\cal Q}\) for the ensemble   \({\cal E}\) is defined as the minimum entropy 
produced when dephased (measured) in a basis from 
\({\cal D}_{\cal A}\).  More precisely,
\[
{\cal Q} ({\cal E}) = \min_{{\cal D}_{\cal A}} \sum_{x=1}^{N} p_x 
S(\sum_{i=1}^{n}p_{i|x}\left|a_i\right\rangle \left\langle a_i \right|).
\]
Here \(p_{i|x}\) is the probability that \(\left|a_i\right\rangle\) clicks if the signal state was 
\(\left|\psi_x\right\rangle\), i.e. \(p_{i|x} = |\left\langle \psi_x | a_i \right\rangle|^2\). 
\(S\) denotes the von Neumann entropy, defined as \(S(\eta) = -\mbox{tr} \eta \log_2 \eta\). 
Since \(\left|a_i\right\rangle\) are orthonormal, 
\(S(\sum_{i=1}^{n}p_{i|x}\left|a_i\right\rangle \left\langle a_i \right|)
= H(\{p_{i|x}\}_{i=1}^{n})\).
 Here \(H\) denotes the Shannon entropy, defined as
\(H(\{q_j\}_{j=1}^{m}) = - \sum_{j=1}^{m} q_j \log_2 q_j\), where 
\(\{q_j\}_{j=1}^{m}\) forms a set of probabilities. Therefore 
\[
{\cal Q} ({\cal E}) = \min_{{\cal D}_{\cal A}} \sum_{x=1}^{N} p_x 
H(\{p_{i|x}\}_{i=1}^{n}).
\]
The minimum is taken over all bases in \({\cal D}_{\cal A}\).

It is interesting to note that 
one may  define 
\({\cal Q}\), for the ensemble \({\cal E}\), as the minimum entropy 
produced when dephased in an \emph{arbitrary} basis in 
\({\cal H}_A \otimes {\cal H}_B\) so that the resulting ensemble after the measurement is an LOCC-distinguishable 
set of states. 


It is quite straightforward to carry over the above definition  
to the case of mixed ensembles.
For the case of mixed states, one must include in the definition, a subtraction of  the initial von Neumann entropy
of the signal states. More precisely, consider an ensemble 
\({\cal E}^{'} = \{p_x, \varrho_x^{AB}\}_{x=1}^{N}\) of bipartite signal 
states, which are 
not necessarily pure. In this case, \({\cal A}\) will denote the union of the supports of \(\varrho_x^{AB}\) 
for \(x= 1, 2, \ldots , N\). The definition of \({\cal D}_{\cal A}\) is then exactly the same as for pure signals.
Then 
\begin{equation}
\label{mixed-er-bela}
{\cal Q}({\cal E}^{'}) =  \min\limits_{{\cal D}_{\cal A}} \sum\limits_{x=1}^{N} p_x 
 \left(H(\{p_{i|x}\}_{i=1}^{n}) - S(\varrho_x)\right),
\end{equation}
where \(p_{i|x} 
= \left\langle a_i \right| \varrho_x^{AB} \left| a_i \right\rangle\).

Actually, the above notion of \({\cal Q}\) as a measure of quantumness can be used in much more general situations
than just in the case of a system consisting of two spatially  localized subsystems. Suppose that a source
produces an ensemble \(\{p_x, \varrho_x\}_{x=1}^{N}\), with the \(\varrho_x\)'s defined on some 
Hilbert space \({\cal H}\). And as before, let \({\cal A}\) denote the union
of the supports of \(\varrho_x\), and let \(\{\left|a_i\right\rangle\}_{i=1}^{n}\) be an arbitrary complete 
basis in \({\cal H}\). Depending on a set of allowed operations, say \(\Lambda\),
 the nonzero overlap \cite{overlap} of  \(\{\left|a_i\right\rangle\}_{i=1}^{n}\)
with \({\cal A}\) may or may not be distinguishable. Let \({\cal D}_{\cal A}^{\Lambda}\) be 
the set of all bases    \(\{\left|a_i\right\rangle\}_{i=1}^{n}\) for which the nonzero overlap 
with \({\cal A}\) is distinguishable under \(\Lambda\). 
Then the measure of quantumness of the ensemble 
\(\Gamma = \{p_x, \varrho_x\}_{x=1}^{N}\), 
with respect to the set of allowed operations \(\Lambda\), 
is defined as 
\begin{equation}
\label{lambda-r-bela}
{\cal Q}(\Gamma)  =  
 \min\limits_{{\cal D}_{\cal A}^\Lambda} \sum\limits_{x=1}^{N} p_x 
 \left(H(\{p_{i|x}\}_{i=1}^{n}) - S(\varrho_x)\right),
\end{equation}
where \(p_{i|x}\) is, as before, the probability that \(\left|a_i\right\rangle\) clicks given that 
the signal was \(\varrho_x\).

When the states \(\varrho_x\) are bipartite states between A and B, with the Hilbert space 
\({\cal H}\) split as 
\({\cal H}_A \otimes {\cal H}_B\), and with the allowed operations being LOCC between A and B, we recover 
the definition given in eq. (\ref{mixed-er-bela}).

For a multiparty ensemble \(\{p_x, \varrho_x^{ABC \ldots}\}\), defined on 
the Hilbert space \({\cal H}_A \otimes {\cal H}_B \otimes {\cal H}_C \otimes \ldots\), 
one chooses the bases \(\{\left|a_i\right\rangle^{ABC \ldots}\}_{i=1}^{n}\) 
spanning this Hilbert space. Then considering the set of allowed operations as 
LOCC between A, B, C, \(\ldots\), \({\cal D}_{\cal A}\) is the set of bases whose
nonzero overlap with \({\cal A}\) is distinguishable by such LOCC operations, where
\({\cal A}\) is the union  of the supports of \(\varrho_x^{ABC \ldots}\).
In this way, we have a measure of quantum correlation of an ensemble of multiparty states.

Importantly, \({\cal Q}\) can be defined also for ensembles of single systems. Then \({\cal Q}\) is 
the quantity in eq. (\ref{lambda-r-bela}), with the allowed operations being all quantum mechanical operations. 
Consequently the dephasing is in any orthonormal basis in \({\cal H}\). This quantity can be interpreted 
as a measure of quantumness of the ensemble \(\Gamma\).




%
%
%
%
%
%
%
%
%

We now show that \({\cal Q}(\Gamma)\) is average relative entropy distance from some ensembles of states. 
We have (see eq. (\ref{lambda-r-bela})) 
\({\cal Q}(\Gamma)  =  \min_{\{\varsigma^{'}_x\}} \sum_x p_x [-\mbox{tr}(\varrho_x \log_2 \varsigma^{'}_x) 
+ \mbox{tr}(\varrho_x \log_2 \varrho_x)] = 
\min\limits_{\{\varsigma^{'}_x\}} \sum_x p_x S(\varrho_x|\varsigma^{'}_x)\), so that
\begin{equation}
\label{tophat}
 {\cal Q}(\Gamma) =  \min\limits_{\{\varsigma_x\}} \sum_x p_x S(\varrho_x|\varsigma_x) ,
\end{equation}
where \(\varsigma^{'}_x = \sum_i p_{i|x} \left| a_i \right\rangle \left\langle a_i \right|\),
\(\varsigma_x = \sum_i a_i^x \left| a_i \right\rangle \left\langle a_i \right|\), 
\(\{\left| a_i \right\rangle\}_i\) being any set of states which are distinguishable under the set of allowable 
operations \(\Lambda\),
 \(\{a_i^x\}_i\)'s being 
arbitrary sets of probabilities, mixing the \(\left| a_i \right\rangle\)'s.
\(S(\varrho|\varsigma) = \mbox{tr}(\varrho \log_2 \varrho - \varrho \log_2 \varsigma)\) is the 
relative entropy distance of \(\varrho\) from \(\varsigma\).
So \({\cal Q}\) is the average relative entropy distance of the signal states from the ``classical'' 
ensembles, namely
the ones whose members are mixtures of distinguishable states (under \(\Lambda\)). Note that for a single 
``classical'' ensemble, the constituent signals are mutually commuting.
Although \({\cal Q}\) turns out to be a relative entropy distance, it has (in contrast to relative 
entropy of entanglement) 
an operational meaning in terms of entropy production (by which it is defined 
in eq. (\ref{mixed-er-bela})) even for a single copy of the ensemble. Therefore \({\cal Q}\) is 
more intimately related to quantum correlation measure of states of \cite{Gdansk} than to relative 
entropy of entanglement.
Note that 
the averaging in eq. (\ref{tophat}) 
is done before the minimization, and hence the measure \({\cal Q}\) can still
``feel'' the ensemble as a whole.

Consider now the case of an ensemble of bipartite states with the allowed operations being LOCC between the 
sharing partners. Suppose also that the union of the supports of the ensemble covers the whole bipartite 
Hilbert space. Then a set of orthogonal states in the support (i.e. a complete orthogonal basis)
 is distinguishable under LOCC only if they are all product states \cite{moren}. 
(Note that the 
opposite is not necessarily true \cite{nlwe}.) Ensembles \(\{\varsigma_x\}\) that are mixtures of states 
from the
sets \(\{\left| a_i \right\rangle\}\) which are distinguishable under LOCC, but do not span the complete 
bipartite Hilbert space, will (in this case) produce infinite relative entropy distance,
 and can therefore be ignored (cf. 
eq. (\ref{tophat})). So the relavant ensembles that appear in formula (\ref{tophat}) are all separable
states. (Again the opposite is not true.) Thus we have 
\begin{equation}
\label{bhalo}
{\cal Q}({\cal E}^{'}) \geq \overline{E_R} = \sum_x p_x E_R(\varrho_x^{AB}),
\end{equation} 
for the ensemble \({\cal E}^{'} = \{p_x, \varrho_x^{AB}\}\), where we have assumed that
the union of the supports of the \(\varrho_x^{AB}\)'s cover the whole
bipartite Hilbert space. \(E_R\) is the relative entropy of 
entanglement defined as \(E_R(\varrho^{AB}) = \min_{\varsigma} S(\varrho|\varsigma)\), where 
the minimization is over all separable states \(\varsigma\) \cite{VPRK}. 
Among other things, the inequality (\ref{bhalo}) 
will be important for
evaluation  of  \({\cal Q}\) for certain bipartite ensembles.

As \({\cal Q}\) is defined in terms of von Neumann entropy, it will inherit some continuity properties 
due to the Fannes' inequality \cite{Fannes}.
By definition, \({\cal Q}\) is vanishing for ensembles which are distinguishable under the allowed set of operations
\(\Lambda\). And it is nonzero otherwise.


It should be noted that \({\cal Q}\), for the case of multipartite ensembles (with the allowed operations 
being the LOCC class) can actually increase under LOCC. This is because one can create
nonorthogonal product states after starting with a multi-orthogonal product basis \cite{eta_holo_multi-orth}.
Note however that \({\cal Q}\) can increase even if the output states are orthogonal, as
  there exist
ensembles of orthogonal \emph{product} states which are locally indistinguishable
 \cite{nlwe}. Such ensembles, which by definition has nonzero \({\cal Q}\), can be 
created by LOCC from a multi-orthogonal product basis \cite{eta_holo_multi-orth}, which being LOCC distinguishable
has by definition vanishing \({\cal Q}\).   
However, from our experience with entanglement-like quantum correlation measures \cite{IBMhuge,
VPRK,Vidal-mon2000,Horodecki_limits,qic}, we know that a quantum correlation measure should 
show some kind of monotonicity.   We believe that \({\cal Q}\), for the case of multipartite ensembles,
will be monotonically decreasing 
under LOCC operations, if we do not allow \emph{tracing out} as a valid operation.     
Note that tracing out can never be useful in a distinguishing protocol. The accessible
information under LOCC, for an ensemble of multipartite states, is the same as the accessible information 
under LOCC without tracing out as a valid operation.  The operations that can be useful in attaining the 
accessible information are 
(a) adding  local ancillas,
(b) local unitarities, 
(c) local dephasing (von Neumann measurement and then forgetting the outcome), 
(d) communication of pre-dephased quantum states (classical communication).
Adding the item (e) tracing out, will give us the whole set of LOCC. \({\cal Q}\) is obviously monotonically
decreasing for operations (a), (b), and (d). We conjecture that 
\({\cal Q}\) is nonincreasing also for local dephasing of multipartite ensembles.
Therefore we have a
natural axiom for any measure of quantum correlation of ensembles of multipartite states, viz. it must be 
nonincreasing under (a), (b), (c), and (d). Let us emphasize that the axiom holds for measures 
of \emph{multipartite} ensembles, not only for bipartite ones. Moreover, nonincreasingness under addition 
of ancillas, unitary operations, and dephasing is in this way seen to be a natural axiom for any measure of 
quantumness of a \emph{single} system.



We will  now calculate the value of our measure for 
the ensemble \({\cal B}\) consisting of the four Bell 
states, 
\(\left|\phi^{\pm} \right\rangle = \frac{1}{\sqrt{2}}(\left|00\right\rangle \pm \left|11\right\rangle)\),    
\(\left|\psi^{\pm} \right\rangle = \frac{1}{\sqrt{2}}(\left|01\right\rangle \pm \left|10\right\rangle)\),
given with equal prior probabilities.
The four Bell states span the whole \(2 \otimes 2\) Hilbert space. To find \({\cal Q}\), we have 
to minimize entropy production after dephasing over all LOCC-distinguishable bases in \(2 \otimes 2\). 
Let us first 
dephase in the computational basis \(\{\left|00\right\rangle, \left|11\right\rangle, \left|01\right\rangle,
\left|10\right\rangle\}\). Suppose for example, that the signal is \(\left|\phi^{+}\right\rangle\).
Then either \(\left|00\right\rangle\) or  \(\left|11\right\rangle\) clicks with equal probabilities. So the 
entropy produced is \(H(\frac{1}{2})\), where \(H(\cdot)\) is the binary entropy function, defined as 
\(H(p)=-p\log_2p-(1-p)\log_2(1-p)\), \(0 \leq p \leq 1\). 
The signal \(\left|\phi^{+}\right\rangle\) is created by the source  with probability \(\frac{1}{4}\). The case is 
similar for the other signals. Therefore
 we obtain 
\({\cal Q} \leq 4 \times \frac{1}{4}H(\frac{1}{2}) = 1\).
However from 
ineq. (\ref{bhalo}) we know that this bound is saturated, as the Bell states have \(E_R =1\). Therefore, we have 
that \({\cal Q} =1\) for the ensemble of the four Bell states given with equal prior probabilities.
Consider now the more general 
ensemble \({\cal B}^{'}\) consisting of 
\(a\left|00\right\rangle + b \left|11\right\rangle\), 
\(\overline{b} \left|00\right\rangle - \overline{a} \left|11\right\rangle\), 
\(c\left|01\right\rangle + d \left|10\right\rangle\), 
\(\overline{d} \left|01\right\rangle - \overline{c} \left|10\right\rangle\),
given with equal prior probabilities.
Again a dephasing in the computational basis 
\(\{\left|00\right\rangle, \left|11\right\rangle, \left|01\right\rangle,
\left|10\right\rangle\}\), implies that 
\({\cal Q} \leq \frac{1}{2}(H(|a|^2) + H(|c|^2))\).
And again saturation follows from ineq. (\ref{bhalo}), so that 
\({\cal Q} = \frac{1}{2}(H(|a|^2) + H(|c|^2))\).
Similar calculations will deliver us also the value of \({\cal Q}\) for  the canonical set of maximally 
entangled states in \(d \otimes d\),  given with equal prior probabilities, viz.
\(\left|  \psi_{nm}^{max}\right\rangle =\frac{1}{\sqrt{d}}\sum_{j=0}^{d-1} e^{2\pi ijn/d}\left|
j\right\rangle \left|  (j+m)\mbox{mod}\quad d\right\rangle\)
(\(n,m=0,\ldots , d-1\)). 
Considering a dephasing of 
this basis
in the computational basis, and using ineq. (\ref{bhalo}), one obtains 
\({\cal Q} = \log_2 d\).

Quantification of quantum correlation of an ensemble of states has 
several important potential applications. E.g., 
it can be used to obtain a complementarity relation with locally accessible information (accessible 
information under LOCC-based measurements). 
For an ensemble \(\{p_x, \varrho_x^{AB}\}\)
of bipartite states, 
a complementarity 
has been obtained between the locally accessible information and \emph{average} shared entanglement of the ensemble 
states \cite{comp}. Precisely, if \(I_{acc}^{LOCC}\) denotes the locally accessible information for the 
ensemble \(\{p_x, \varrho_x^{AB}\}\), then one has \cite{comp}
\begin{equation}
\label{comp-ek}
I_{acc}^{LOCC} + \overline{E} \leq \log_2 n,
\end{equation}
where ensemble states \(\varrho^{AB}_x\) are defined on \({\cal H}_A \otimes {\cal H}_B\), 
\(n = \dim \{{\cal H}_A \otimes {\cal H}_B\}\), and \(\overline{E} = \sum_{x} p_x E(\varrho^{AB}_x)\),
 \(E\) being any asymptotic entanglement measure of bipartite states \cite{qic}.

However the complexity of locally accessible information in comparison to 
globally accessible information depends on the geometry of the ensemble, which cannot be captured by taking an 
\emph{average} of an individual property (for example, entanglement) of the ensemble states. Consequently, the 
complementarity obtained in \cite{comp} can potentially 
be made stronger if the locally accessible information is taken along with 
a measure of quantum correlation of an ensemble of states.

We conjecture that the following complementarity relation holds between the accessible information, 
\(I_{acc}^{\Lambda}\),   for 
the set of allowed operations \(\Lambda\), and 
\({\cal Q}\), for an ensemble \(\{p_x,\varrho_x\}_{x=1}^{N}\):
\begin{equation}
\label{comp-dui1}
I_{acc}^{\Lambda} + {\cal Q} \leq  \log_2 N.
\end{equation}
Among other things, note also the change in the right-hand-side of ineq. (\ref{comp-dui1}) with respect to 
that of ineq. (\ref{comp-ek}). In ineq. (\ref{comp-ek}), the right-hand-side is the logarithm of 
the dimension of the Hilbert space on which the signals are defined. However in ineq. (\ref{comp-dui1}), the 
right-hand-side is the logarithm of the number of states in the ensemble, which can be less than, equal to, or 
greater than the dimension of the said Hilbert space.

For an ensemble of bipartite states, \(\{p_x,\varrho_x^{AB}\}_{x=1}^{N}\), the proposed complementarity is 
\begin{equation}
\label{comp-dui}
I_{acc}^{LOCC} + {\cal Q} \leq  \log_2 N.
\end{equation}
Note that due to ineq. (\ref{bhalo}), the relation (\ref{comp-dui}) will in general be stronger than the 
one in ineq. (\ref{comp-ek}), if the ensemble consists of \(n = \dim \{{\cal H}_A \otimes {\cal H}_B\}\) states
that spans \({\cal H}_A \otimes {\cal H}_B\), and if we replace \(E\) by \(E_R\). We conjecture that the relation 
(\ref{comp-dui}) is in general stronger than the one in (\ref{comp-ek}), at least when we replace \(E\) by 
\(E_R\). In particular, we believe that \({\cal Q} \geq \overline{E_R} + \log_2 N - \log_2 n\).

For the case of the four Bell states, given with equal prior probabilities, \(I_{acc}^{LOCC} = 1\). This follows 
from ineq. (\ref{comp-ek}) and the fact that measuring in the computational basis gives \(I_{acc}^{LOCC} \geq 1\).
And for this case, we have proven that \({\cal Q} = 1\). Therefore we have proven inequality (\ref{comp-dui})
for the case of four Bell states given with equal probabilities.
Ineq. (\ref{comp-dui}) is true also for the more general ensemble \({\cal B}^{'}\).
From ineq. (\ref{comp-ek}), we have that for the ensemble \({\cal B}^{'}\),
\(
I_{acc}^{LOCC} \leq 2 - \frac{1}{2}(H(|a|^2) + H(|c|^2))
\).
But a measurement in the computational basis shows that this bound can be achieved. 
Using the value of \({\cal Q}\) obtained for \({\cal B}^{'}\),
we have the inequality (\ref{comp-dui}) proven for this ensemble.
One may similarly prove the inequality (\ref{comp-dui}) for the canonical set of maximally entangled states in 
\(d \otimes d\).

For the case of the ensemble of the three Bell states \(\left|\phi^{\pm}\right\rangle\),
\(\left|\psi^{+}\right\rangle\), given with equal prior probabilities,
a measurement in the 
computational basis gives \(I_{acc}^{LOCC} \geq \log_2 3 - \frac{2}{3}\). 
And the fact that set \(\{\left|00\right\rangle, \left|11\right\rangle, \left|\psi^{+}\right\rangle\}\) is 
an LOCC-distinguishable ensemble in the span of 
\(\left|\phi^{\pm}\right\rangle\),
\(\left|\psi^{+}\right\rangle\), gives
\({\cal Q} \leq \frac{2}{3}\).  
Note here that these two inequalities do not have a contradiction with the proposed ineq. (\ref{comp-dui}).

One may consider ineq. (\ref{comp-dui1}) for the case of an ensemble of a single system, with all quantum mechanically 
allowed operations in \(\Lambda\). Then ineq. (\ref{comp-dui1}) can be seen as a 
``Heisenberg uncertainty relation''. The accessible information is the maximal ``information gain'' that is 
possible about the system. On the other hand, \({\cal Q}\) denotes the quantumness of the ensemble, quantifying 
the resistance to such information gain.
 In this case of a single system, it is actually possible to prove the ineq. (\ref{comp-dui1})
in a restricted case.
For an ensemble \(\{p_x, \varrho_x\}_{x=1}^{N}\), 
defined on the Hilbert space \({\cal H}\),
the accessible information, 
\(I_{acc} = \max_{M}\left(H(\{p_x\}_{x=1}^{N}) - \sum_{y} r_y H(\{p_{x|y}\}_{x=1}^{N}) \right)\),
where a measurement \(M=\{y\}\) has been performed, the measurement result \(y\) having occured with 
probability \(r_y\). \(p_{x|y}\) is the probability that the signal was \(\varrho_x\), given that 
the outcome was \(y\). The maximization is carried over all measurements. The quantity within the 
maximization is the mutual information between the source producing the ensemble \(\{p_x, \varrho_x\}_{x=1}^{N}\)
and the maesurement outcomes. 
But we know that \(H(X) - H(X|Y) = H(Y) - H(Y|X)\), for random variables \(X\) and \(Y\). Therefore 
\(H(\{p_x\}_{x=1}^{N}) - \sum_{y} r_y H(\{p_{x|y}\}_{x=1}^{N}) 
= 
H(\{p_y\}_{y}) - \sum_{x=1}^{N} p_x H(\{p_{y|x}\}_{y})\).
At this point, we restrict ourselves to only projection valued measurements on \({\cal H}\). Then we have 
\(I_{acc}  \leq  \max_{M} H(\{p_y\}_{y}) - {\cal Q} 
          \leq   \log_2 \dim {\cal H} - {\cal Q} 
         \leq   \log_2 N - {\cal Q}\), 
which is just the inequality in (\ref{comp-dui1}).

It is conceivable that the complementarity in ineq. (\ref{comp-dui1}) is 
true only when we consider a certain form of regularisation of the quantity \({\cal Q}\).
Given an ensemble \(\{p_x, \varrho_x\}\) and a set of allowed operations \(\Lambda\), consider 
\({\cal Q}_0\) for an extension
\(\{p_x, \varrho_x \otimes \left|0\right\rangle \left\langle 0 \right|\}\) of the given ensemble, where 
\(\left|0\right\rangle\) is any state that is free under \(\Lambda\). The regularization is then the 
minimum of \({\cal Q}_0\)'s for all possible such extensions.
This brings in the possibility of vanishing regularized \({\cal Q}\) even for some 
ensembles that are indistinguishable under \(\Lambda\). 


This work is supported by
EU grants RESQ and QUPRODIS, and by  the University of Gda\'{n}sk, 
Grant No. BW/5400-5-0256-3.


\begin{thebibliography}{10}
\bibitem{overlap} 
The vector \(\left|a_{i_k}\right\rangle\) of the basis 
\(\{\left|a_i\right\rangle^{AB}\}_{i=1}^{n}\) has a ``nonzero overlap'' with \({\cal A}\), if 
\(\left\langle a_{i_k} | \wp \right\rangle \ne 0\), for some vector \(\left| \wp \right\rangle \in {\cal A}\). 
The set of such \(\left|a_{i_k}\right\rangle\) from the basis 
\(\{\left|a_i\right\rangle^{AB}\}_{i=1}^{n}\) is called the ``overlap of the basis 
\(\{\left|a_i\right\rangle^{AB}\}_{i=1}^{n}\) with \({\cal A}\)'' or  the ``part of the basis
\(\{\left|a_i\right\rangle^{AB}\}_{i=1}^{n}\) having nonzero overlap with \({\cal A}\)''.
\bibitem{eta_holo_multi-orth}
For  orthogonal bases 
\(\{\left|i\right\rangle\}_{i=1}^{\dim {\cal H}_k}\)  of   \({\cal H}_k\) 
\((k= A, B, \ldots)\),   
\(\{\left|i\right\rangle \otimes \left|j\right\rangle \otimes 
\ldots \}_{i,j,\ldots=1}^{\dim {\cal H}_A,\dim {\cal H}_B, \ldots}\) 
  is the corresponding multi-orthogonal basis. Such a basis is trivially distinguishable by LOCC between 
A, B, \(\ldots\). So an ensemble whose states are from such a 
basis, will have 
zero \({\cal Q}\).
\end{thebibliography}
\end{document}